\documentclass[twocolumn,A4]{article}
\usepackage[dvips]{graphics}
\usepackage{setspace}
\usepackage{color}
\definecolor{gold}{rgb}{0.85,0.66,0}
\definecolor{dblue}{rgb}{0,0,0.8}
\begin{document}
\onecolumn
\begin{center}
{\bf{\Large {\textcolor{gold}{NOT gate response in a mesoscopic ring: 
An exact result}}}}\\
~\\
{\textcolor{dblue}{Santanu K. Maiti}}$^{1,2,*}$ \\
~\\
{\em $^1$Theoretical Condensed Matter Physics Division,
Saha Institute of Nuclear Physics, \\
1/AF, Bidhannagar, Kolkata-700 064, India \\
$^2$Department of Physics, Narasinha Dutt College,
129, Belilious Road, Howrah-711 101, India} \\
~\\
{\bf Abstract}
\end{center}
We explore NOT gate response in a mesoscopic ring threaded by a magnetic 
flux $\phi$. The ring is attached symmetrically to two semi-infinite 
one-dimensional metallic electrodes and a gate voltage, viz, $V_a$, is
applied in one arm of the ring which is treated as the input of the NOT 
gate. The calculations are based on the tight-binding model and the Green's 
function method, which numerically compute the conductance-energy and 
current-voltage characteristics as functions of the ring-to-electrodes 
coupling strength, magnetic flux and gate voltage. Our theoretical study 
shows that, for $\phi=\phi_0/2$ ($\phi_0=ch/e$, the elementary flux-quantum) 
a high output current ($1$) (in the logical sense) appears if the input 
to the gate is low ($0$), while a low output current ($0$) appears when 
the input to the gate is high ($1$). It clearly exhibits the NOT gate 
behavior and this aspect may be utilized in designing an electronic 
logic gate. 

\vskip 1cm
\begin{flushleft}
\end{flushleft}
\vskip 4.5in
\noindent
{\bf ~$^*$Corresponding Author}: Santanu K. Maiti

Electronic mail: santanu.maiti@saha.ac.in
\newpage
\twocolumn

\section{Introduction}

The study of electron transport in quantum confined geometries has
become one of the most fascinating branch of nanoscience and
technology. With the help of different advanced technologies, the
simple looking quantum confined systems can be used in designing 
nanodevices especially in electronic as well as spintronic engineering. 
The idea of manufacturing nanodevices are based on the concept of
quantum interference effect which is generally vanishes for larger
systems. On the other hand, for much smaller sizes the quantum phase
coherence is maintained across the sample. A mesoscopic metallic ring 
is one such promising example where electronic motion is confined and
the transport becomes predominantly coherent. Using a mesoscopic ring 
we can make a device that can act as a logic gate, which may be used 
in nanoelectronic circuits. To explore this phenomenon we design a 
bridge system where the ring is sandwiched between two external 
electrodes, so-called the electrode-ring-electrode bridge. The ring 
is then subjected to an Aharonov-Bohm (AB) flux $\phi$ which is the key 
controlling factor for the whole logical operation in this particular 
geometry. The theoretical description of electron transport in a bridge 
system has got much progress following the pioneering work of Aviram and 
Ratner.$^1$ Later, many excellent experiments$^{2-4}$ have been done in 
several bridge systems to understand the basic mechanisms underlying the 
electron transport. Though extensive studies on electron transport have 
already been done both theoretically$^{5-13}$ as well as 
experimentally,$^{2-4}$ yet lot 
of controversies are still present between the theory and experiment, 
and the complete knowledge of the conduction mechanism in this scale 
is not very well established even today. For illustrative purposes, 
here we mention some of these issues as follow. The electronic transport
in the ring changes drastically depending on the interface geometry 
between the ring and the electrodes. By changing the geometry, one 
can tune the transmission probability of an electron across the ring 
which is solely due to the effect of quantum interference among the 
electronic waves passing through different arms of the ring. Not only
that, the electron transport in the ring can be modulated in other 
way by tuning the magnetic flux, that threads the ring. The AB flux 
threading the ring may change the phases of the wave functions 
propagating along the different arms of the ring leading to constructive 
or destructive interferences, and therefore, the transmission amplitude 
changes.$^{14-18}$ Beside these factors, ring-to-electrodes coupling is 
another important issue that controls the electron transport in a 
meaningful way.$^{18}$ All these are the key factors which regulate the 
electron transmission in the electrode-ring-electrode bridge system and 
these effects have to be taken into account properly to reveal the 
transport mechanisms. 

Our main aim of the present work is to study the NOT gate response in a 
mesoscopic ring threaded by a magnetic flux $\phi$. The ring is contacted 
symmetrically to the electrodes, and a gate voltage $V_a$ is applied in 
one arm of the ring (see Fig.~\ref{not}) which is regarded as the input 
of the NOT gate. A simple tight-binding model is used to describe the 
system and all the calculations are done numerically. Here we address the 
NOT gate behavior by studying the conductance-energy and current-voltage 
characteristics as functions of the ring-electrodes coupling strength, 
magnetic flux and gate voltage. Our study reveals that for a particular 
value of the magnetic flux, $\phi=\phi_0/2$, a high output current ($1$) 
(in the logical sense) is available if the input to the gate is low ($0$), 
while if the input to the gate is high ($1$), a low output current ($0$) 
appears. This phenomenon clearly shows the NOT gate behavior. To the best 
of our knowledge the NOT gate response in such a simple system has yet 
not been addressed in the literature.

The scheme of the paper is as follow. Following the introduction 
(Section $1$), in Section $2$, we describe the model and the 
theoretical formulations for the calculation. Section $3$ explores 
the results, and finally, we conclude our study in Section $4$.

\section{Model and the synopsis of the theoretical background}

\begin{figure}[ht]
{\centering \resizebox*{6.5cm}{5cm}{\includegraphics{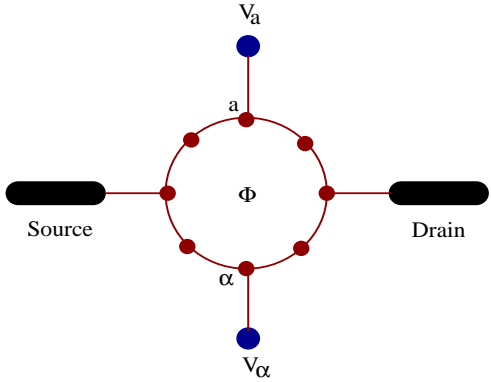}}\par}
\caption{Schematic representation for the operation of a NOT gate.
The atomic sites $a$ and $\alpha$ are subjected to the voltages 
$V_a$ and $V_{\alpha}$ respectively, those are variable (for color 
illustration, see the web version).}
\label{not}
\end{figure}

Let us start with the model presented in Fig.~\ref{not}. A mesoscopic ring, 
subjected to an AB flux $\phi$, is attached symmetrically (upper and lower 
arms have equal number of lattice points) to two semi-infinite 
one-dimensional ($1$D) metallic electrodes. A gate voltage $V_a$, taken 
as the input voltage of the NOT gate, is applied to the atomic site $a$ in 
the upper arm of the ring. While, an additional gate voltage $V_{\alpha}$ 
is applied to the site $\alpha$ in the lower arm of the ring. Both these 
two voltages are variable.

At very low temperature and bias voltage the conductance $g$ of the ring 
can be expressed from the Landauer conductance formula,$^{19-20}$
\begin{equation}
g=\frac{2e^2}{h} T
\label{equ1}
\end{equation}
where $T$ gives the transmission probability of an electron across 
the ring. This $(T)$ can be represented in terms of the Green's 
function of the ring and its coupling to the two electrodes by the 
relation,$^{19-20}$
\begin{equation}
T={\mbox{Tr}} \left[\Gamma_S G_{R}^r \Gamma_D G_{R}^a\right]
\label{equ2}
\end{equation}
where $G_{R}^r$ and $G_{R}^a$ are respectively the retarded and 
advanced Green's functions of the ring including the effects of 
the electrodes. Here $\Gamma_S$ and $\Gamma_D$ describe the 
coupling of the ring to the source and drain respectively. For 
the complete system i.e., the ring, source and drain, the Green's 
function is defined as,
\begin{equation}
G=\left(E-H\right)^{-1}
\label{equ3}
\end{equation}
where $E$ is the injecting energy of the source electron. To Evaluate
this Green's function, the inversion of an infinite matrix is needed since
the complete system consists of the finite ring and the two semi-infinite 
electrodes. However, the entire system can be partitioned into sub-matrices 
corresponding to the individual sub-systems and the Green's function for 
the ring can be effectively written as,
\begin{equation}
G_{R}=\left(E-H_{R}-\Sigma_S-\Sigma_D\right)^{-1}
\label{equ4}
\end{equation}
where $H_{R}$ is the Hamiltonian of the ring that can be expressed within 
the non-interacting picture like,
\begin{eqnarray}
H_{R} & = & \sum_i \left(\epsilon_{i0} + V_a \delta_{ia} + V_{\alpha} 
\delta_{i\alpha}
\right) c_i^{\dagger} c_i \nonumber \\
 & + & \sum_{<ij>} t \left(c_i^{\dagger} c_j e^{i\theta}+ c_j^{\dagger}
c_i e^{-i\theta}\right)
\label{equ5}
\end{eqnarray}
In this Hamiltonian $\epsilon_{i0}$'s are the site energies for all the 
sites $i$ except the sites $i=a$ and $\alpha$ where the gate voltages 
$V_a$ and $V_{\alpha}$ are applied, those are variable. These gate 
voltages can be incorporated through the site energies as expressed in 
the above Hamiltonian. $c_i^{\dagger}$ ($c_i$) is the creation (annihilation) 
operator of an electron at the site $i$ and $t$ is the nearest-neighbor 
hopping integral. The phase factor $\theta=2 \pi \phi/N \phi_0$ comes 
due to the flux $\phi$ threaded by the ring, where $N$ corresponds to 
the total number of atomic sites in the ring. Similar kind of 
tight-binding Hamiltonian is also used, except the phase factor 
$\theta$, to describe the $1$D perfect electrodes where the 
Hamiltonian is parametrized by constant on-site potential $\epsilon_0$ 
and nearest-neighbor hopping integral $t_0$. The hopping integral between
the source and the ring is $\tau_S$, while it is $\tau_D$ between the
ring and the drain. The parameters $\Sigma_S$ and $\Sigma_D$ 
in Eq.~(\ref{equ4}) represent the self-energies due to the coupling of 
the ring to the source and drain respectively, where all the informations 
of this coupling are included into these self-energies.

The current passing through the ring is depicted as a single-electron
scattering process between the two reservoirs of charge carriers. The
current $I$ can be computed as a function of the applied bias voltage 
$V$ by the expression,$^{19}$
\begin{equation}
I(V)=\frac{e}{\pi \hbar}\int \limits_{E_F-eV/2}^{E_F+eV/2} T(E)~ dE
\label{equ8}
\end{equation}
where $E_F$ is the equilibrium Fermi energy. Here we assume that the 
entire voltage is dropped across the ring-electrode interfaces, and it 
is examined that under such an assumption the $I$-$V$ characteristics 
do not change their qualitative features. 

All the results in this communication are determined at absolute
zero temperature, but they should valid even for finite temperature
($\sim 300$ K), since the broadening of the energy levels of the ring 
due to its coupling with the electrodes becomes much larger than that of
the thermal broadening.$^{19}$ For simplicity, we take the unit 
$c=e=h=1$ in our present calculation. 

\section{Results and discussion}

To discuss the results, first we mention the values of the different 
parameters those are used for the numerical calculation. The on-site 
energy $\epsilon_{i0}$ of the ring is taken as $0$ for all the sites 
$i$, except the sites $i=a$ and $\alpha$ where the site energies are 
taken as $V_a$ and $V_{\alpha}$ respectively, and the nearest-neighbor 
hopping strength $t$ is set to $3$. On the other hand, for the side 
attached electrodes the on-site energy ($\epsilon_0$) and the 
nearest-neighbor hopping strength ($t_0$) are fixed to $0$ and $4$ 
respectively. The voltage $V_{\alpha}$ is set to $2$. Throughout the 
\begin{figure}[ht]
{\centering \resizebox*{7cm}{8.5cm}{\includegraphics{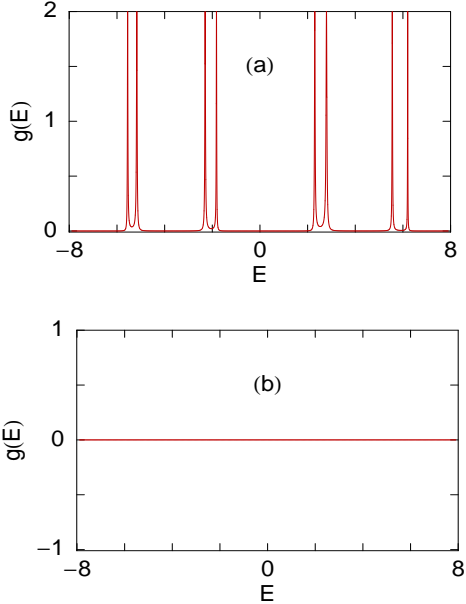}}\par}
\caption{$g$-$E$ curves in the weak-coupling limit for a mesoscopic 
ring with $N=8$, $V_{\alpha}=2$ and $\phi=0.5$. (a) $V_a=0$ and 
(b) $V_a=2$ (for color illustration, see the web version).}
\label{condlow}
\end{figure}
study, we focus our results for the two limiting cases depending on the 
strength of the coupling of the ring to the source and drain. In one 
case we use the condition $\tau_{S(D)} << t$, which is so-called the 
weak-coupling limit. For this regime we choose $\tau_S=\tau_D=0.5$. 
In the other case the condition $\tau_{S(D)} \sim t$ is used, which 
is named as the strong-coupling limit. In this particular regime, the 
values of the parameters are set as $\tau_S=\tau_D=2.5$. The significant 
parameter for all these calculations is the magnetic flux $\phi$ which 
is set to $\phi_0/2$ i.e., 0.5 in our chosen unit.

In Fig.~\ref{condlow} we show the variation of the conductance ($g$) as
a function of the injecting electron energy ($E$), in the limit of
weak-coupling, for a mesoscopic ring with $N=8$ and $V_{\alpha}=2$.
Figures~\ref{condlow}(a) and (b) correspond to the results for the
input voltages $V_a=0$ and $V_a=2$ respectively. For the particular
case when the input voltage $V_a=2$ i.e., the input is high, the
conductance $g$ vanishes (Fig.~\ref{condlow}(b)) in the complete 
energy range. This indicates that the conduction of the electron 
from the source to drain through the ring is not possible. The
situation becomes completely different for the case when the input to 
the gate is zero ($V_a=0$). The result is shown in Fig.~\ref{condlow}(a),
where the conductance shows sharp resonance peaks for some fixed
\begin{figure}[ht]
{\centering \resizebox*{7cm}{8.5cm}{\includegraphics{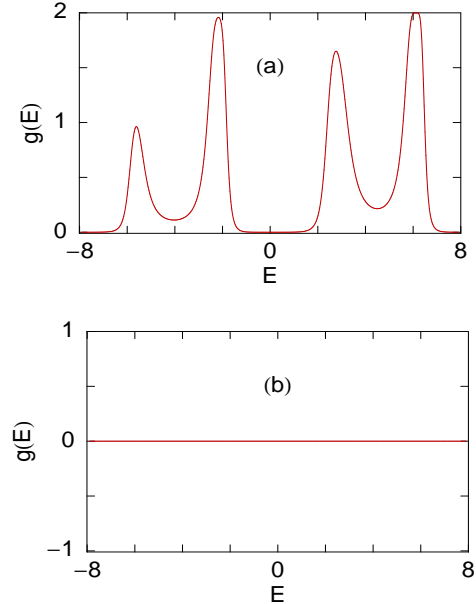}}\par}
\caption{$g$-$E$ curves in the strong-coupling limit for a mesoscopic 
ring with $N=8$, $V_{\alpha}=2$ and $\phi=0.5$. (a) $V_a=0$ and 
(b) $V_a=2$ (for color illustration, see the web version).}
\label{condhigh}
\end{figure}
energies. This reveals the electron conduction across the ring. At
these resonance $g$ approaches the value $2$, and therefore, the
transmission probability $T$ becomes unity, since the expression $g=2T$ 
is satisfied from the Landauer conductance formula (see Eq.~(\ref{equ1}) 
with $e=h=1$). These resonance peaks are associated with the energy 
eigenvalues of the ring, and thus, it can be predicted that the 
conductance spectrum manifests itself the electronic structure of the 
ring. Hence, more resonance peaks are expected for the larger rings, 
associated with their energy spectra. Now we focus the dependences of
the gate voltages on the electron transport for the two different 
cases of the input voltage. The transmission probability of getting
an electron through the ring depends on the quantum interference of
the electronic waves passing through the two arms (upper and lower) 
of the ring. For the symmetrically connected ring i.e., when the two 
arms of the ring are identical with each other, the probability amplitude 
is exactly zero ($T=0$) for the flux $\phi=\phi_0/2$. This is due to 
the result of the quantum interference among the two waves in the two 
arms of the ring, which can be established by a very simple mathematical 
calculation. Therefore, for the case when the input to the gate is
equal to $2$ i.e., $V_a=2$, the upper and lower arms of the ring become 
exactly similar. This is because the potential $V_{\alpha}$ is also 
set to $2$. Accordingly, the transmission probability drops to zero.
If the input voltage $V_a$ is different from the potential applied in
\begin{figure}[ht]
{\centering \resizebox*{7cm}{8.5cm}{\includegraphics{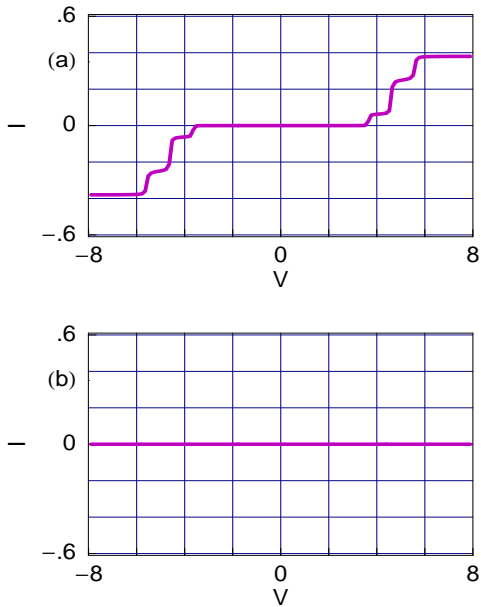}}\par}
\caption{$I$-$V$ curves in the weak-coupling limit for a mesoscopic 
ring with $N=8$, $V_{\alpha}=2$ and $\phi=0.5$. (a) $V_a=0$ and 
(b) $V_a=2$ (for color illustration, see the web version).}
\label{currlow}
\end{figure}
the atomic site $\alpha$, then the two arms are not identical with each
other and the transmission probability will not vanish. Thus, to get 
the zero transmission probability when the input is high, we should 
tune $V_{\alpha}$ properly, observing the input potential and vice versa.
On the other hand, due to the breaking of the symmetry of the two arms,
the non-zero value of the transmission probability is achieved in the
particular case where the input voltage is zero ($V_a=0$), which reveals 
the electron conduction across the ring. From these results we can
emphasize that the electron conduction takes place through the ring 
if the input to the gate is zero, while if the input is high, the 
conduction is no longer possible. This aspect clearly describe the NOT 
gate behavior. Following with this, here we also describe the effect
of the ring-to-electrodes coupling. As representative examples, in 
Fig.~\ref{condhigh} we plot the conductance-energy characteristics for 
the strong-coupling limit, where (a) and (b) are drawn respectively for 
the same input voltages as in Fig.~\ref{condlow}. In the strong-coupling 
case, all the resonances get substantial widths compared to the 
weak-coupling case. This is due to the broadening of the energy 
levels of the ring in the limit of strong coupling, where the 
contribution comes from the imaginary parts of the self-energies 
$\Sigma_S$ and $\Sigma_D$ respectively.$^{19}$ Therefore, by 
tuning the coupling strength, we can get the electron transmission 
across the ring for the wider range of energies and it provides an 
important behavior in the study of current-voltage ($I$-$V$) 
characteristics.

All these features of electron transport become much more clearly
illustrated from our presented current-voltage ($I$-$V$) characteristics.
\begin{table}[ht]
\begin{center}
\caption{NOT gate behavior in the limit of weak-coupling. The current 
$I$ is computed at the bias voltage $6.02$.}
\label{table1}
~\\
\begin{tabular}{|c|c|}
\hline \hline
Input ($V_a$) & Current ($I$) \\ \hline 
$0$ & $0.378$ \\ \hline
$2$ & $0$ \\ \hline \hline
\end{tabular}
\end{center}
\end{table}
The current across the ring is determined by integrating the transmission 
function $T$ as prescribed in Eq.~(\ref{equ8}). The transmission function 
varies exactly similar to that of the conductance spectrum, differ only 
in magnitude by the factor $2$ since the relation $g=2T$ holds from the 
Landauer conductance formula Eq.~(\ref{equ1}).
As illustrative examples, in Fig.~\ref{currlow} we plot the $I$-$V$ 
characteristics for a mesoscopic ring with $N=8$ and $V_{\alpha}=2$, 
in the limit of weak-coupling. For $V_a=0$, the result is shown in
Fig.~\ref{currlow}(a), while Fig.~\ref{currlow}(b) gives the result for
$V_a=2$. Clearly we see that, when the input to the gate is identical 
to $2$ (high), the current $I$ becomes exactly zero (see 
Fig.~\ref{currlow}(b)) for the entire bias voltage $V$. This feature 
is understood from the conductance spectrum, Fig.~\ref{condlow}(b), 
since the current is computed from the integration method of the 
transmission function $T$. On the other hand, a non-zero value of the
current is obtained when the input voltage $V_a=0$, as given in 
Fig.~\ref{currlow}(a). The figure shows that the current exhibits 
staircase-like structure with fine steps as a function of the applied 
bias voltage. This is due to the existence of the fine resonance peaks 
in the conductance spectrum in the weak-coupling limit. With the increase 
of the bias voltage $V$, the electrochemical potentials on the electrodes 
are shifted gradually, and finally cross one of the quantized energy 
\begin{figure}[ht]
{\centering \resizebox*{7cm}{8.5cm}{\includegraphics{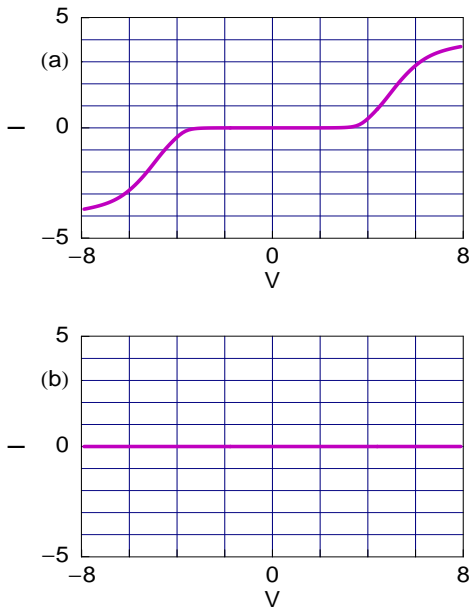}}\par}
\caption{$I$-$V$ curves in the strong-coupling limit for a mesoscopic 
ring with $N=8$, $V_{\alpha}=2$ and $\phi=0.5$. (a) $V_a=0$ and 
(b) $V_a=2$ (for color illustration, see the web version).}
\label{currhigh}
\end{figure}
levels of the ring. Accordingly, a current channel is opened up which 
provides a jump in the $I$-$V$ characteristic curve. Here, it is also 
important to note that the non-zero value of the current appears beyond a
\begin{table}[ht]
\begin{center}
\caption{NOT gate behavior in the limit of strong-coupling. The current 
$I$ is computed at the bias voltage $6.02$.}
\label{table2}
~\\
\begin{tabular}{|c|c|}
\hline \hline
Input ($V_a$) & Current ($I$) \\ \hline 
$0$ & $2.846$ \\ \hline
$2$ & $0$ \\ \hline \hline
\end{tabular}
\end{center}
\end{table}
finite value of the bias voltage $V$, so-called the threshold voltage 
($V_{th}$). This $V_{th}$ can be changed by controlling the size ($N$) 
of the ring. From these current-voltage characteristic curves, the NOT 
gate behavior of the ring can be observed very nicely.
To make it much clear, in Table~\ref{table1}, we show a quantitative
estimate of the typical current amplitude determined at the bias
voltage $V=6.02$. It shows that, when the input to the gate is zero,
the current gets the value $0.378$. While, the current becomes exactly
zero when the input voltage $V_a=2$. In the same fashion, as above, here 
we also describe the $I$-$V$ characteristics for the limit of
strong-coupling. In this limit, the current varies almost continuously 
with the applied bias voltage and gets much larger amplitude than the 
weak-coupling case as presented in Fig.~\ref{currhigh}. The fact is that, 
in the limit of strong-coupling all the energy levels get broadened 
which provide larger current in the integration procedure of the 
transmission function $T$. Thus by tuning the strength of the 
ring-to-electrodes coupling, we can achieve very large current amplitude 
from the very low one for the same bias voltage $V$. All the other 
properties i.e., the dependences of the gate voltages on the $I$-$V$ 
characteristics are exactly similar to those as given in Fig.~\ref{currlow}.
In this strong-coupling limit we also make a quantitative study for the
typical current amplitude, given in Table~\ref{table2}, where the current
amplitude is determined at the same bias voltage ($V=6.02$) as
earlier. The response of the output current is exactly similar to that as
given in Table~\ref{table1}. Here the current achieves the value $2.846$ 
when the input to the gate is zero ($V_a=0$), while it ($I$) becomes 
exactly zero for the case where $V_a=2$. The non-zero value of the current 
in this strong-coupling limit is much larger than the weak-coupling case, 
as expected. From these results we can clearly manifest that a mesoscopic
ring exhibits the NOT gate response.

\section{Concluding remarks}

In this presentation, we have addressed the NOT gate response in a 
mesoscopic metallic ring threaded with a magnetic flux $\phi$ in the Green's
function formalism. The ring is attached symmetrically to the electrodes
and a gate voltages $V_a$ is applied in one arm of the ring which is taken 
as the input of the NOT gate. A simple tight-binding model is used to 
describe the system and all the calculations are done numerically. We have 
computed the conductance-energy and current-voltage characteristics as 
functions of the gate voltage, ring-to-electrodes coupling strength and 
magnetic flux. Very interestingly we have observed that, for the half 
flux-quantum value of $\phi$ ($\phi=\phi_0/2$), a high output current 
($1$) (in the logical sense) appears if the input to the gate is low ($0$). 
While, if the input to the gate is high ($1$), a low output current ($0$) 
appears. It clearly manifests the NOT gate behavior and this aspect may be 
utilized in designing a tailor made electronic logic gate. 

The importance of this article is particularly concerned with (i) the 
simplicity of the geometry and (ii) the smallness of the size. To the 
best of our knowledge the NOT gate response in such a simple 
low-dimensional system that can be operated even at finite temperature 
($\sim 300$ K) has not been addressed earlier in the literature.


\begin{thebibliography}{99}

\bibitem{aviram} A. Aviram and M. Ratner, Chem. Phys. Lett. 29, 277
(\textbf{1974}).
\bibitem{reed1} J. Chen, M. A. Reed, A. M. Rawlett, and J. M. Tour, Science
286, 1550 (\textbf{1999}).
\bibitem{reed2} M. A. Reed, C. Zhou, C. J. Muller, T. P. Burgin, and J. M.
Tour, Science 278, 252 (\textbf{1997}).
\bibitem{tali} T. Dadosh, Y. Gordin, R. Krahne, I. Khivrich, D. Mahalu,
V. Frydman, J. Sperling, A. Yacoby, and I. Bar-Joseph, Nature 436,
677 (\textbf{2005}).
\bibitem{orella1} P. A. Orellana, M. L. Ladron de Guevara, M. Pacheco,
and A. Latge, Phys. Rev. B 68, 195321 (\textbf{2003}).
\bibitem{orella2} P. A. Orellana, F. Dominguez-Adame, I. Gomez, and
M. L. Ladron de Guevara, Phys. Rev. B 67, 085321 (\textbf{2003}).
\bibitem{nitzan1} A. Nitzan, Annu. Rev. Phys. Chem. 52, 681 (\textbf{2001}).
\bibitem{nitzan2} A. Nitzan and M. A. Ratner, Science 300, 1384
(\textbf{2003}).
\bibitem{muj1} V. Mujica, M. Kemp, and M. A. Ratner, J. Chem. Phys.
101, 6849 (\textbf{1994}).
\bibitem{muj2} V. Mujica, M. Kemp, A. E. Roitberg, and M. A. Ratner,
J. Chem. Phys. 104, 7296 (\textbf{1996}).
\bibitem{walc2} K. Walczak, Phys. Stat. Sol. (b) 241, 2555 (\textbf{2004}).
\bibitem{walc3} K. Walczak, arXiv:0309666.
\bibitem{cui} W. Y. Cui, S. Z. Wu, G. Jin, X. Zhao, and Y. Q. Ma,
Eur. Phys. J. B. 59, 47 (\textbf{2007}).
\bibitem{baer2} R. Baer and D. Neuhauser, J. Am. Chem. Soc. 124,
4200 (\textbf{2002}).
\bibitem{baer3} D. Walter, D. Neuhauser, and R. Baer, Chem. Phys.
299, 139 (\textbf{2004}).
\bibitem{tagami} K. Tagami, L. Wang, and M. Tsukada, Nano Lett. 4,
209 (\textbf{2004}).
\bibitem{walc1} K. Walczak, Cent. Eur. J. Chem. 2, 524 (\textbf{2004}).
\bibitem{baer1} R. Baer and D. Neuhauser, Chem. Phys. 281, 353
(\textbf{2002}).
\bibitem{datta} S. Datta, {\em Electronic transport in mesoscopic systems},
Cambridge University Press, Cambridge (\textbf{1997}).
\bibitem{marc} M. B. Nardelli, Phys. Rev. B 60, 7828 (\textbf{1999}).

\end{thebibliography}
\end{document}